\documentclass[aps,prb,preprint,unsortedaddress]{revtex4-1}
\usepackage{amssymb,mathbbol,amsbsy}
\usepackage{amsfonts}
\usepackage{amsmath}
\usepackage{subfigure, epsfig}
\usepackage{graphics, graphicx}

\newcommand{\bra}[1]{\langle #1|}
\newcommand{\ket}[1]{|#1\rangle}
\newcommand{\braket}[2]{\langle #1|#2\rangle}

\begin{document}

\title{Dynamical bi-stability of single-molecule junctions: A combined experimental/theoretical study of PTCDA  on Ag(111)}

\author{T. Brumme}
\affiliation{Institute for Materials Science and Max Bergmann Center of Biomaterials, Dresden University of Technology, 01062 Dresden, Germany}
\altaffiliation{These authors contributed equally to this work}

\author{O. A. Neucheva}
\affiliation{Peter Gr\"unberg Institut (PGI-3) and JARA---Fundamentals of Future Information Technology, Forschungszentrum J\"ulich, 52425 J\"ulich, Germany}
\affiliation{Institute of Materials Research and Engineering, 3 Research Link, Singapore 117602, Republic of Singapore}
\altaffiliation{These authors contributed equally to this work}

\author{C. Toher}
\author{R. Guti{\'e}rrez}
\affiliation{Institute for Materials Science and Max Bergmann Center of Biomaterials, Dresden University of Technology, 01062 Dresden, Germany}

\author{C. Weiss}
\author{R. Temirov}
\affiliation{Peter Gr\"unberg Institut (PGI-3) and JARA---Fundamentals of Future Information Technology, Forschungszentrum J\"ulich, 52425 J\"ulich, Germany}

\author{A. Greuling}
\author{M. Kaczmarski}
\author{M. Rohlfing}
\affiliation{Fachbereich Physik, Universit\"at Osnabr\"uck, 49069 Osnabr\"uck, Germany}

\author{F. S. Tautz}
\affiliation{Peter Gr\"unberg Institut (PGI-3) and JARA---Fundamentals of Future Information Technology, Forschungszentrum J\"ulich, 52425 J\"ulich, Germany}

\author{G. Cuniberti}
\affiliation{Institute for Materials Science and Max Bergmann Center of Biomaterials, Dresden University of Technology, 01062 Dresden, Germany}
\affiliation{National Center for Nanomaterials Technology, POSTECH, Pohang 790-784, Republic of Korea}
\email{research@nano.tu-dresden.de}

\begin{abstract}
The dynamics of a molecular junction consisting of a PTCDA molecule between the tip of a scanning tunneling microsope and a Ag(111) surface have been investigated
experimentally and theoretically. Repeated switching of a PTCDA molecule between two conductance states is studied by low-temperature scanning tunneling microscopy
for the first time, and is found to be dependent on the tip-substrate distance and the applied bias. Using a minimal model Hamiltonian approach combined with
density-functional calculations, the switching is shown to be related to the scattering of electrons tunneling through the junction, which progressively excite the
relevant chemical bond. Depending on the direction in which the molecule switches, different molecular orbitals are shown to dominate the transport and thus the
vibrational heating process. This in turn can dramatically affect the switching rate, leading to non-monotonic behavior with respect to bias under certain
conditions. In this work, rather than simply assuming a constant density of states as in previous works, it was modeled by Lorentzians. This allows for
the successful description of this non-monotonic behavior of the switching rate, thus demonstrating the importance of modeling the density of states realistically.
\end{abstract}


\maketitle

\section{introduction}
The scanning tunneling microscope (STM) is a valuable and versatile tool for the study and manipulation of nanoscale structures.\cite{binning, moresco} In scanning
mode, it can be used to image surfaces with atomic resolution, and to probe the electronic density of states at a range of energy values. Alternatively, it can be
brought into contact with surface features to form junctions and measure transport properties.\cite{Tao,PTCDA_Kondo,Pump1,Grill,Lindsay,Toher1} Nanostructures and devices can
be manipulated and fabricated using an STM, with the possibility to pick up and deposit atoms and molecules using the tip.\cite{eigler,eigler2,stroscio, Lindsay}
An important aspect related to the tip-molecule interaction is the telegraph noise observed in the conductance in certain circumstances, which originates from the
repeated switching of single atoms or functional groups between different stable
configurations.\cite{stroscio,PhysRevLett.86.672,stroscio2,Tao,VioletaIancu09122006,werner,PeterLiljeroth08312007,Kern}
Several physical mechanisms have been proposed to explain this phenomenon:
thermal activation, vibrational heating (for intermediate biases)\cite{Gao, ueba2004, ueba2005, ueba2009, domcke2004} and transition through an electronic excited
state with no conformational bi-stability (for high biases).\cite{Elste} If the masses involved are not too large (i.e. for a single atom), quantum tunneling is
also possible.\cite{Lauhon} 

\begin{figure}
 \centering
   \subfigure[]{\includegraphics[height=4.5cm]{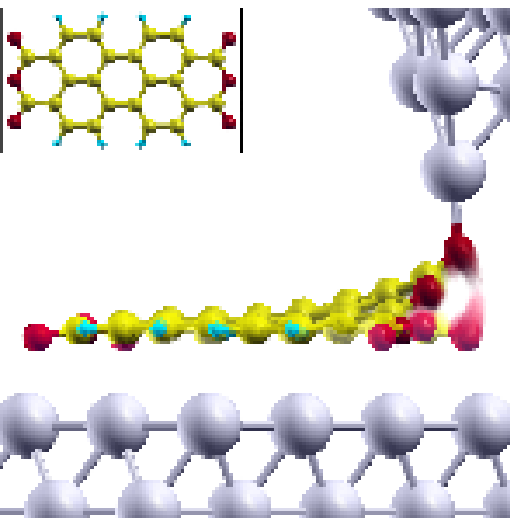}}
   \subfigure[]{\includegraphics[height=4.5cm,clip]{./figure_1b.eps}}
   \subfigure[]{\includegraphics[height=4.5cm]{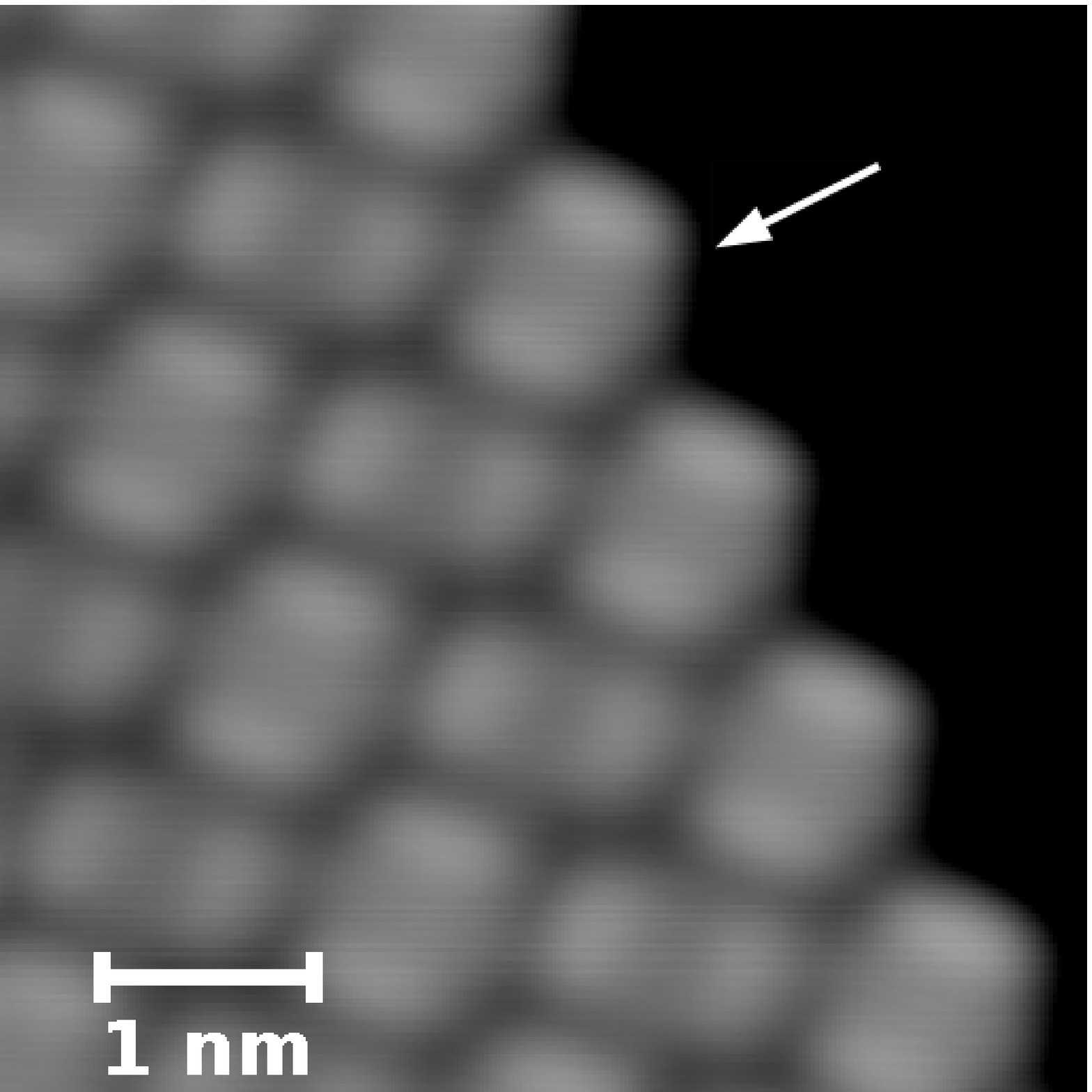}}
 \caption{(color online) (a) Schematic picture illustrating the up- and down-states and the switching between the two for the tip-PTCDA-Ag(111) junction
 (red--oxygen, yellow--carbon, light-blue--hydrogen, grey--silver).
 The atomic coordinates are taken from the DFT calculations described in Ref.~\citenum{Toher1}. The inset shows the structure of the gas phase molecule.
 (b) Measured current at 125mV during approach of the STM tip above the carboxylic oxygen of the PTCDA molecule in (c).
 The tip was moved by 0.6\AA{} at a rate of 1\AA{} per 23 min. (c) STM image of the edge of a monolayer of PTCDA. The white arrow indicates the PTCDA molecule
 which was used for the switching measurements, and points to the oxygen atom which interacts with the tip.}
\label{figure_1}
\end{figure}

In this work, we present a systematic study of this switching behavior in the specific system of perylene-3,4,9,10-tetracarboxylic-dianhydride (PTCDA, inset in
Fig.~\ref{figure_1}(a)) on Ag(111), using both experimental and theoretical methods. With results from density-functional (DFT) calculations and by extending a
microscopic model developed in Ref.~\citenum{Gao} to describe the coupling of an adsorbate energy level to the adsorbate vibrational excitations, a good agreement
with the experimentally measured switching rates can be achieved. 

PTCDA deposited on Ag(111) forms a highly ordered metal-organic interface, the electronic and geometric structure of which has been well-characterized using a
variety of both experimental and theoretical techniques.\cite{Gloeckler,tautz07,tautz07a, Kraft} The PTCDA molecules form long-range ordered commensurate monolayers on the
Ag(111) substrate with two flat-lying chemisorbed molecules per unit cell in a herringbone arrangement (see Ref.~\citenum{Gloeckler}). The chemisorption results in
the former lowest unoccupied molecular orbital (LUMO) of the isolated molecule being shifted below the Fermi level of the silver surface, so that there is charge
transfer from the substrate to the molecule thus producing a net negative charge on the molecule.\cite{Kraft}

In previous experiments we have found that it is possible to form a chemical bond between the carboxylic oxygen atoms and the STM tip, if the latter is approached
towards the molecule above one of the carboxylic oxygen atoms:\cite{PTCDA_Kondo,Pump1,Toher1} the oxygen atom, followed by part of the carbon skeleton of the PTCDA
molecule, jumps into contact with the tip. The most likely distance for this
single switch to happen (without applying a bias voltage)
is 6.65\AA{}.\cite{PTCDA_Kondo} In a theoretical analysis, carried out by
calculating potential profiles of relaxed PTCDA molecules between tip and surface as a function of oxygen-surface separation for a range tip sample separations,
we found the spontaneous jump into contact at 6.2\AA{},\cite{Toher1} in good agreement with experiment.

Once the molecular junction with the tip has been formed, there are two possible ways for the molecule to behave when the tip is retracted: either the molecule is
peeled off from the surface completely or it falls back to the surface.\cite{PTCDA_Kondo} We have further observed that, under certain conditions (see below), the
current fluctuates in time between a high- and a low-conductance state, see e.g. Fig.~\ref{figure_1}(b) in which the telegraph noise in the current is evident.
These two-state fluctuations can be explained by the switching of the molecule in and out of contact with the tip (see Fig.~\ref{figure_1}(a)). In the
high-conductance state, one of the carboxylic oxygen atoms of the molecule forms a chemical bond with the tip (``up-state''), establishing a two-terminal molecular
junction, while in the low-conductance state the molecule is bonded exclusively to the surface (``down-state'') so that a tunnel barrier is now present between the
tip and the molecule. These switching processes of the molecule can also be seen in the topographic images taken with the tip very close to the surface.\cite{PTCDA_Kondo}

\section{experimental methods}
Our experiments have been performed with a CREATEC low temperature scanning tunneling microscope (5-6 K) in ultrahigh vacuum with a base pressure below
10$^{-10}$mbar. The Ag(111) surface has been prepared by repeated sputtering/annealing cycles (Ar$^+$ ion energy 0.8 keV, annealing at approximately 850 K).
Surface quality has been controlled in situ with low-energy electron diffraction (LEED). The PTCDA molecules have been evaporated from a Knudsen cell at 580 K
onto the surface at room temperature. An electrochemically etched tungsten wire has been used as the STM tip, which has been cleaned in situ by annealing. The
final atomic sharpening has been done by the indentation of the tip into the clean metal substrate and/or by the application of voltage pulses. Tip quality has
been checked by measuring the surface state of Ag(111). The PTCDA material (commercial purity 99\%) has been purified by resublimation and outgassing in ultra 
high vacuum. 

\begin{figure}
 \centering
 \subfigure[]{\includegraphics[width=0.45\textwidth,clip]{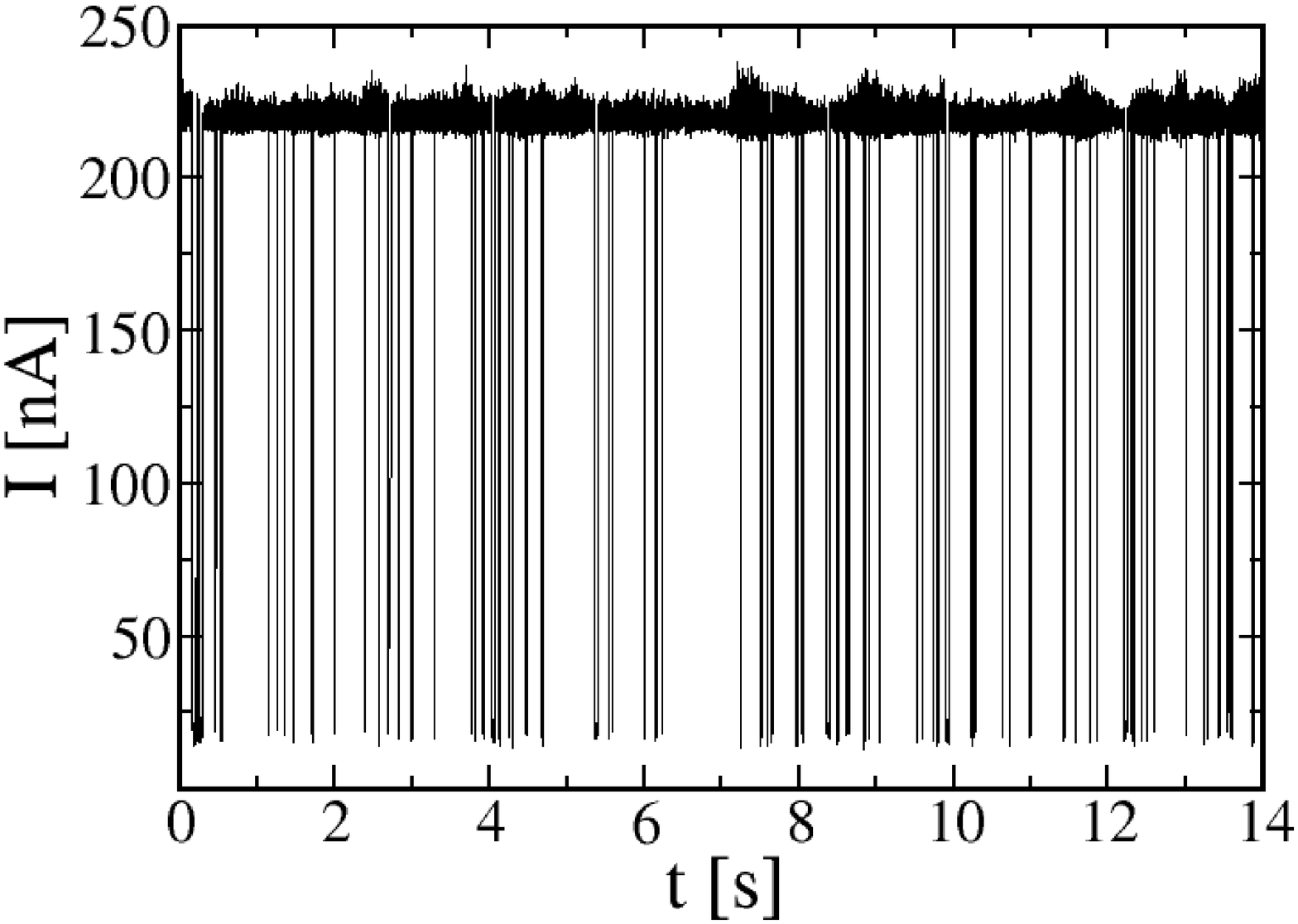}} 
 \subfigure[]{\includegraphics[width=0.45\textwidth]{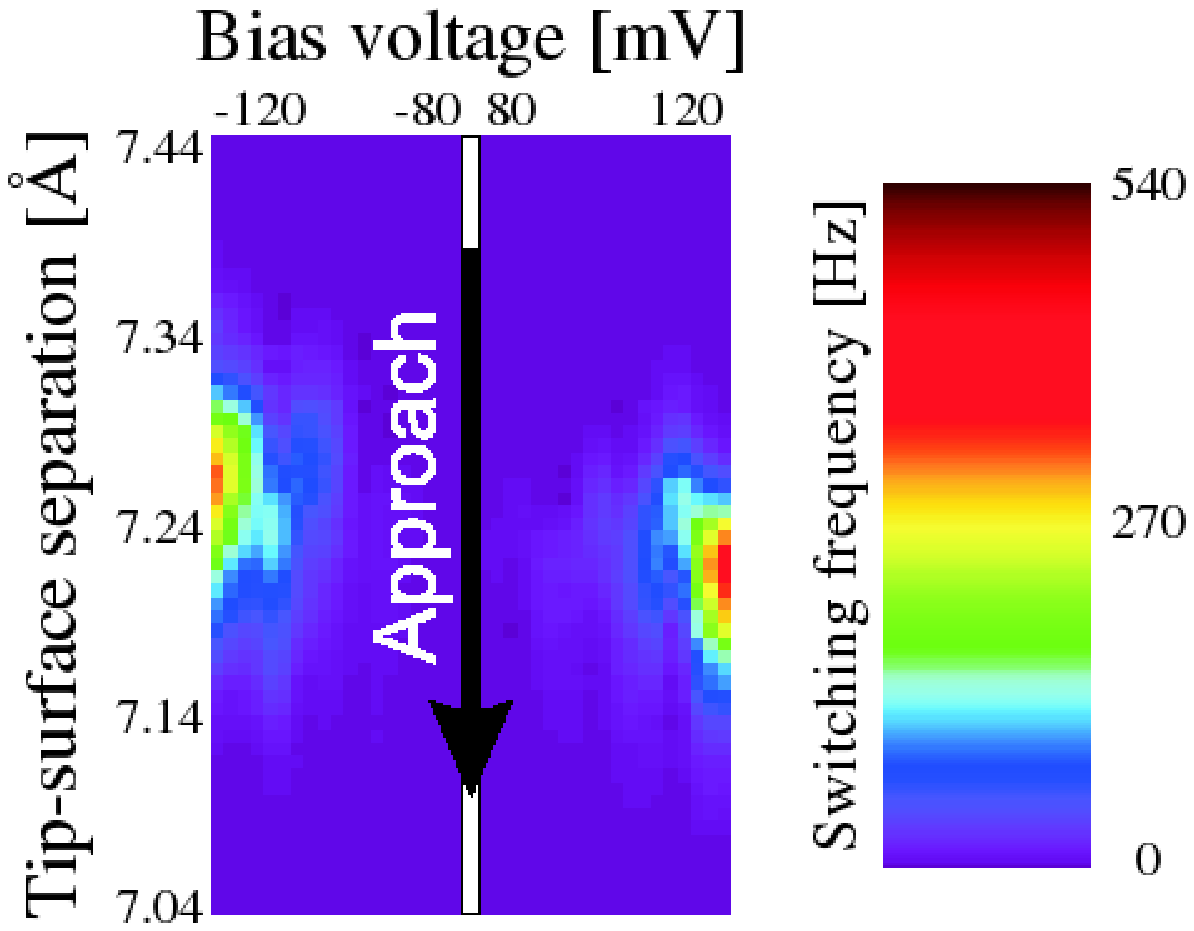}}
 \subfigure[]{\includegraphics[width=0.45\textwidth,clip]{./figure_2c.eps}}
 \subfigure[]{\includegraphics[width=0.45\textwidth,clip]{./figure_2d.eps}}
 \caption{(color online) Switching of a PTCDA molecule between up- and down- state. 
 (a) Current vs. time trace measured 
 at 95mV with a tip-surface separation of 7.1\AA{}. (b) Map of the average switching frequency as
 function of bias voltage and tip-surface separation. The corresponding spectra were measured at constant bias during tip approach. 
 The bias range from -120mV to 120mV was covered with a step of 5mV. (c), (d) Residence time histograms for the up-state and the down-state,
 extracted from the time trace in panel (a). The red solid lines show the exponential fit used to extract the transfer rate $R$.}
 \label{experiments}
\end{figure}

Prior to the measurement of the switching process, the STM tip was stabilized at $V_{bias}$=-340mV and $I$=0.1nA, corresponding to a tip-surface separation of
10.6\AA{} (Ref.~\citenum{Toher1}), which is outside the regime in which repeated switching is observed. Absolute calibration of the tip-surface separation was
done as described in Ref.~\citenum{Toher1} (error of $\pm0.5$\AA{} for the absolute height).
Time spectra of the current were recorded for different bias voltages and tip-surface separations with the feed-back
loop switched off. The time dependent current I(t) is shown in Fig.~\ref{experiments}(a) for the applied bias voltage of 95mV and with the tip positioned at
7.1\AA{} above the substrate.

The quantitative analysis of the switching process, which is the primary objective of this paper, has been carried out for molecules located at the edge of a
monolayer island of PTCDA/Ag(111) (as indicated with the white arrow in in Fig.~\ref{figure_1}(b)).
The reason for choosing these molecules is that the PTCDA molecules in the midst of a compact layer are more difficult to
pick up due to strong intermolecular interactions with neighboring molecules via hydrogen bonds,\cite{WeissJACS} while isolated molecules do not always fall
back to the same position on the surface when they switch from the up- to the down-state, thereby leaving the junction and precluding the continued measurement
of the switching time trace.

A color-coded map of the frequency of switching events as a function of bias voltage and tip-surface separation is displayed in Fig.~\ref{experiments}(b). We
observe the following: (1) Repeated switching occurs for both bias voltage polarities above a threshold of approximately $\vert 100\vert$meV. In contrast, for
$U_{bias}<\vert 100\vert$meV a single jump into contact occurs\cite{PTCDA_Kondo,Toher1} (not indicated in Fig.~\ref{experiments}(b)); for these bias voltages,
the junction may only (but does not necessarily) switch back from the up- to the down-state if the tip is retracted again beyond the tip-surface separation at
which the jump into contact has originally occured (hysteresis). (2) Repeated switching occurs in a narrow bracket of tip-surface separations in the range
from 7.34\AA{} to 7.14\AA{}. (3) The range in which repeated switching is observed appears at slightly larger tip-surface separations for negative bias than
for positive bias. This latter fact may be related to the negative polarization of the carboxylic oxgen atoms in Ag(111)-adsorbed PTCDA.     

\begin{figure}[bth]
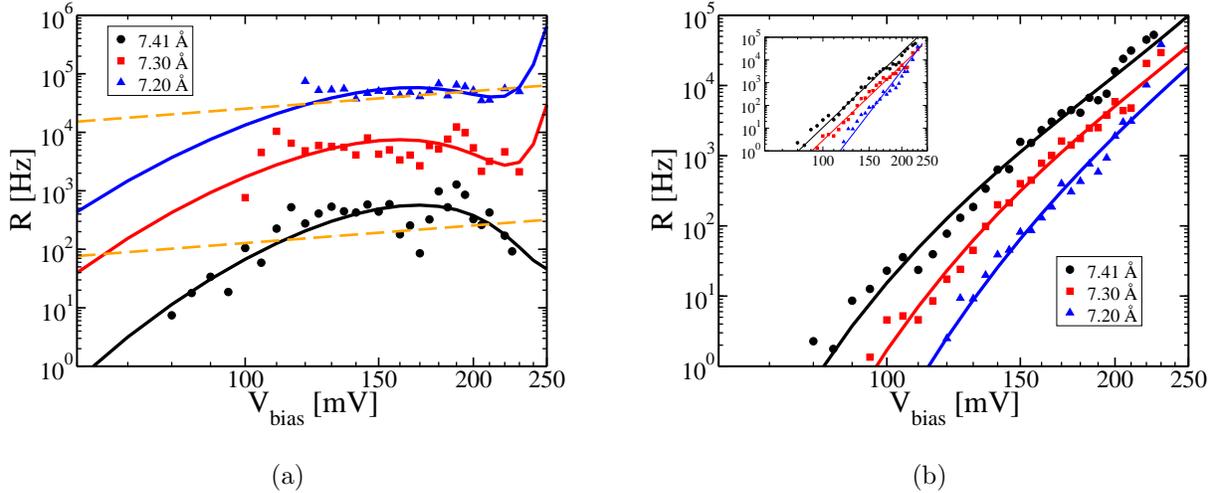

 \subfigure[]{\includegraphics[width=0.45\linewidth,clip]{./figure_3a.eps}}\hspace{1cm}
 \subfigure[]{\includegraphics[width=0.45\linewidth,clip]{./figure_3b.eps}}
 \caption{(color online) Double logarithmic plot of the transfer rate for PTCDA switching between the STM tip and a Ag(111)
surface for different tip-surface separations. Measured transfer rates for
switches from (a) surface to tip and (b) tip to surface are indicated by small symbols. Solid
lines display the theoretical transfer rate, (a) dashed orange lines the model of Ref.~\citenum{Gao}. The inset in (b) shows a possible fit with
the model of Ref.~\citenum{Gao}. However, the parameters thus obtained disagree with both experiment and DFT calculations.}
 \label{fit_pioa}
\end{figure}

>From the I(t) curve in Fig.~\ref{experiments}(a) one can see that for the chosen bias and tip-surface separation, the up-state is preferred: the statistical
residence time analysis (Figs.~\ref{experiments}(c) and (d)) reveals a difference of more than one order of magnitude in the residence time values for the
high- and low-conductance states. The single exponential behavior of the curves indicates a two state Markovian switching process where the residence time
probability density $P$ is given by the expression $P\left(t\right)\:=\:R\,\exp\left(-\,R\,t\right).$ Here, $R$ is the transfer rate between the two
conductance states. It is obtained by fitting the equation for $P$ to the corresponding residence time histogram. By performing such a transfer rate analysis
for different bias voltages one can determine the transfer rates as a function of bias for every measured tip-surface separation. The rates for three typical
tip-surface separations are displayed in Fig.~\ref{fit_pioa}. The tip$\rightarrow$surface transfer rate increases monotonically with applied bias in the given
voltage range, but the surface$\rightarrow$tip transfer rate appears to have a maximum around 180mV. Finally, both rates are dependent on the tip-surface
separation, as can also be seen in the experimental data in Fig.~\ref{experiments}(b).

\section{theoretical model}
\begin{figure}
 \subfigure[]{\includegraphics[width=0.5\textwidth,clip]{./figure_4a.eps}}
  \subfigure[]{\includegraphics[width=0.45\textwidth]{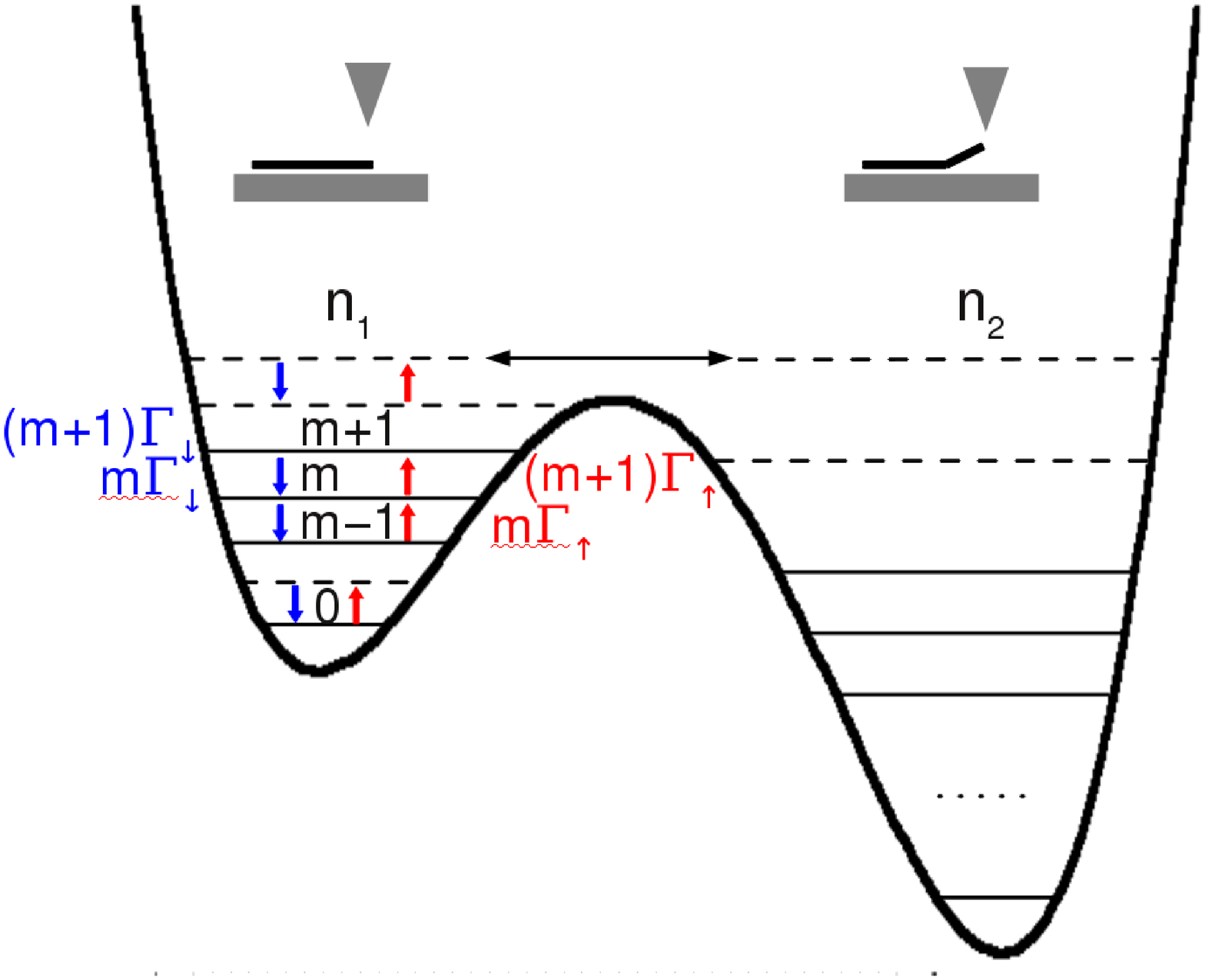}}
 \caption{(color online) (a) Double well potentials of relaxed PTCDA molecules between tip
 and surface as a function of oxygen-surface separation for a 
 range tip-sample separations calculated with DFT for a single PTCDA molecule.\cite{Toher1}
 (b) Schematic double well potential used to describe up- and down-states of PTCDA in the STM junction, including vibrational levels and model
 parameters appearing in Eq.~(\ref{escape_rate_1}). $\Gamma_{\downarrow/\uparrow}$ denote the relaxation 
 and excitation rate of a molecular vibration due to scattering of tunneling electrons, and $n_1$ and $n_2$ denote the critical number of vibrations which 
 have to be excited to induce the switching.}
 \label{fig_overview}
\end{figure}

To gain insight into the observed current switching, we first focus on the nature of the coupling between PTCDA and the surface, and then provide a link to
the experimental data by applying a model calculation. The mechanisms of the chemical bonding of PTCDA to Ag(111) includes  hybridization of the molecular
orbitals with the substrate states, charge transfer between the substrate and the molecule, local bonds of the carboxylic oxygens to silver atoms below and an
extended bond of the molecular $\pi$-system  to the surface.\cite{Pump1,tautz07,tautz07a,Kraft} Assuming that the two meta-stable positions can be well-represented by a (not
necessarily symmetric) double-well potential, the transfer of an adsorbate between the two minima may involve a variety of physical processes, such as
(i) thermal activation, (ii) quantum tunneling, (iii) a transition through an electronic excited state with no conformational bi-stability, or (iv) vibrational
heating. Process (i) is of minor interest in this work, since the experiments are performed at very low temperatures (5-6 K) and the barrier height is larger
than 100meV, which excludes the thermal activation. Due to the relatively large mass of the part of the molecule involved in the switching process, process (ii)
is also very improbable. Assuming a tunneling barrier of 100meV height (measured from the vibrational ground state) and 1\AA{} width
(cf. Fig.~\ref{fig_overview}(a)) the corresponding tunneling rate for the carboxylic oxygen atom was estimated to be of the order of $10^{-8}$Hz. For
process (iii), which involves an excited state of the molecule, the residence time of the tunneling electrons has to be sufficiently large to induce this
excitation. However, since the molecule is chemisorbed on the Ag(111) surface, this residence time is expected to be quite small, so that process (iii) also
seems unlikely in this case. Thus, we suggest that the microscopic mechanism leading to switching is related to  vibrational heating, where the transition is
induced by progressive vibrational excitation of the relevant chemical bond (i.e., either the oxygen-surface bond for the surface$\rightarrow$tip process or the
oxygen-tip bond for the reverse process) by the inelastic scattering of tunneling electrons, eventually leading to bond breaking. The transition rate is then
mainly determined by the competition between energy gain from the tunneling charges and energy losses due to electron-hole pair generation and/or coupling to the
substrate phonon continuum. 

In Ref.~\citenum{Gao} Gao et al.~ developed a theoretical model to describe atomic switching by vibrational heating. They concluded that the switching rate should
exhibit a power law dependence $R\propto {V_{bias}}^n$ on the bias voltage $V_{bias}$ where $n$ is the number of vibrational levels that have to be climbed before
the switch can occur. In our experiments we observe a striking difference between the tip$\rightarrow$surface and surface$\rightarrow$tip switching processes as
far as the bias dependence of the switching rate is concerned (cf. Fig.~\ref{fit_pioa}). While the tip$\rightarrow$surface process shows an almost linear
$R(V_{bias})$ behavior in the double logarithmic plot, in essential agreement with the prediction of Ref.~\citenum{Gao}, a reasonable description within the model
of Gao \textit{et al.}\cite{Gao} for the surface$\rightarrow$tip process is very unlikely, because $R(V_{bias})$ deviates from a simple power law, showing a
saturation of the transfer rate at approximately 120mV, with even a possible maximum around 180mV. Moreover, a (rather poor) fit of the data in
Fig.~\ref{fit_pioa}(a) with $R\propto {V_{bias}}^n$ would only be possible by assuming $n=1$ (cf. dashed line in Fig.~\ref{fit_pioa}(a)), which is in contrast to
the calculated potential energy surfaces, see Fig.~\ref{fig_overview}(a).

Below we show that if the energy dependence of the density of states around the Fermi level is taken into account explicitly, and if in particular
\textit{different} transport orbitals for the two configurations (i.e. up- and down-states) are used, the evident differences between the surface$\rightarrow$tip
and tip$\rightarrow$surface processes can be rationalized and both the data in Figs.~\ref{fit_pioa}(a) and (b) can be fitted with parameters which are in 
qualitative agreement with DFT results for the electronic structure of the molecular junction. In  our model, we will neglect the coupling to the substrate phonon
continuum, since the anharmonic coupling is, in general, very small at low temperatures. 

Our model is a minimal approach based on that used in Ref.~\citenum{Gao} to describe the vibrational heating. The Hamiltonian describing the tunneling of
electrons between the STM tip and the surface via an adsorbate level (in this case, the adsorbate being the PTCDA molecule) has the following form:
\begin{align}
\label{full_hamiltonian}
 H =&\sum_s\:\varepsilon_s\,c_s^\dagger\,c_s\:+\:\sum_t\:\varepsilon_t\,c_t^\dagger\,c_t\:+\:\varepsilon_m\,c_m^\dagger\,c_m\:+\:\hbar\,\omega\,b^\dagger\,b\\
    &\:+\:\sum_s\:\left( T_{sm}\,c_s^\dagger\,c_m\:+\:H.c. \right)\:+\:\sum_t\:\left(T_{tm}\,c_t^\dagger\,c_m\:+\:H.c. \right)\notag\quad.
\end{align}
Here $s$, $t$ and $m$ label one-electron states $\ket s$, $\ket t$ and $\ket m$ of the surface, the tip and the molecule, respectively, with the
corresponding energies $\varepsilon_s$, $\varepsilon_t$ and $\varepsilon_m$. The hopping between the surface and the tip via the molecular level
is described by the two terms including $T_{sm}$ and $T_{tm}$. The
coupling between the vibrational motion of the molecule and the electron propagating through it can be modeled by:
\begin{align}
\label{h_ev}
    H_{e-v} =&\lambda_0\:\left(b^\dagger\:+\:b\right)\,\left(c_m^\dagger\,c_m\right)
\quad,
\end{align}
where,
\begin{align}
\label{lambda0}
  \lambda_0 =&\sqrt{\frac{\hbar}{2\,M\,\omega}}\:\varepsilon_m'\quad.
\end{align}
The coupling is modeled by assuming that $\varepsilon_m$ is a linear function of the vibrational coordinate $q$, $\varepsilon_m(q)$; $\omega$ is the frequency of
the molecular vibration with the normal coordinate $q=\sqrt{\frac{\hbar}{(2M\omega)}}\,(b^\dagger+b)$ and mass $M$,
and $\varepsilon_m'=\partial\varepsilon_m/\partial q$ at $q=0$.

Since the effect of the electron-vibration interaction on the adsorbate electronic states is in general weak, it can be treated by first-order perturbation theory.
The assumed linearity in the charge-vibron coupling simplifies the problem since only the excitation and relaxation rates,
$\Gamma_\uparrow$ and $\Gamma_\downarrow$, between the vibrational ground state and the first excited state are required (cf. Fig.~\ref{fig_overview}(b)). In
first-order perturbation theory these transition rates are given by Fermi's Golden Rule:
\begin{align}
\label{rate_up}
 \Gamma_\uparrow =&2\frac{2\pi}{\hbar}\:\sum_{j,l}\left|\bra{j,1}H_{e-v}\ket{l,0}\right|^2\:f_l\left(1-f_j\right)\:\delta(\varepsilon_j-\varepsilon_l+\hbar\omega),\\
\label{rate_down}
 \Gamma_\downarrow =&2\frac{2\pi}{\hbar}\:\sum_{j,l}\left|\bra{j,0}H_{e-v}\ket{l,1}\right|^2\:f_l\left(1-f_j\right)\:\delta(\varepsilon_j-\varepsilon_l-\hbar\omega)
\end{align}
where $0$ and $1$ are the vibrational ground state and the first excited state respectively, while $j$ and $l$ denote any of the stationary one-electron states of
the tip or the substrate with corresponding Fermi-Dirac distributions $f_{j,l}=1/\left\{1+exp\left[(\varepsilon-\varepsilon_{l,j})/(k_B\,T)\right]\right\}$, and
$H_{e-v}$ denotes the electron-vibration interaction (Eq.~(\ref{h_ev})).

These rates describe the vibrational excitation and relaxation induced by the tunneling electrons. Since the initial and final states of a tunneling electron can
be located either in the tip or the substrate, these rates can be decomposed into four different terms: $\Gamma_{\uparrow,\downarrow}^{ss}$,
$\Gamma_{\uparrow,\downarrow}^{tt}$, $\Gamma_{\uparrow,\downarrow}^{st}$ and $\Gamma_{\uparrow,\downarrow}^{ts}$, which sum up to give $\Gamma_{\uparrow,\downarrow}$.
Here, the first (second) superscript denotes whether the final (initial) state belongs to the surface or the tip. In contrast to Ref.~\citenum{Gao}, we will not
assume that the adsorbate local DOS is constant over the relevant energy range, but rather we model it by a Lorentzian shape,
$\rho^{s,t}_{m}(E)=\Delta_{s,t}/((E-\varepsilon_{m})^{2}+\Delta^{2})$, where $\Delta=\Delta_s+\Delta_t$, with $\Delta_s$ and $\Delta_t$ describing the coupling
between the molecular level and the substrate and tip electronic states, respectively. Using this function, the excitation and relaxation rates can be calculated
analytically in the low temperature limit. We refer the interested reader to appendix A for further details and a comprehensive description of the calculation.

To describe the transfer between the two possible meta-stable states a truncated harmonic oscillator model, as described in Ref.~\citenum{Gao}, is adopted. The
transfer rate $R$ can be expressed as a product of the transition into level $n$ (see Fig.~\ref{fig_overview}(b)) and an effective Boltzmann factor
(with characteristic temperature $T_\nu=\hbar\omega/(k_B\ln[\Gamma_\downarrow/\Gamma_\uparrow])$) describing the
probability to arrive at the sub-critical level $n-1$ where the transition takes place:\cite{Gao}
\begin{align}
\label{escape_rate_1}
 R \simeq&\:n\,\Gamma_\uparrow\,\exp\left[\frac{(n-1)\,\hbar\omega}{k_B\,T_\nu}\right]\:=\:n\,\Gamma_\uparrow\:\left(\frac{\Gamma_\uparrow}{\Gamma_\downarrow}\right)^{n-1}\quad.
\end{align}
Since the adsorbate local DOS is not assumed to be constant over the relevant energy range, the above expression in general does \textit{not} yield a simple power
law dependence on the applied bias as in Ref.~\citenum{Gao} ($R\propto {V_{bias}}^n$). This simple scaling law can only be recovered, if the molecular level is
situated far from the Fermi energy (so that the DOS at $\varepsilon _F$ is almost constant).

\section{results and discussion}
Using Eqs.~(A1--A4) in the appendix, we are now able to fit the transfer rate in Eq.~(\ref{escape_rate_1}) to the experimental results. Figure \ref{fit_pioa}
shows the fitted transfer rates as a function of bias voltage, together with the experimental data. The corresponding fitting parameters are listed in
Tab.~\ref{table_full} and will now be discussed in detail.

The vibrational energies $\hbar\omega$ ({\it i.e.} the size of the steps on the ``vibrational ladder'') were determined from the curvature of the calculated
potential energy surfaces, shown in Fig.~\ref{fig_overview}(a). They lie around 19meV for the shallower well of the down-state, and around 40meV for the deeper
well of the up-state. The dependence of these vibrational frequencies on the tip--surface distance is negligible (cf. Tab.~\ref{table_full}).

The $n$ are an output of the fitting of the transfer rates. Multiplied with $\hbar\omega$, they yield the barrier heights for the switching process. The products
$n_1\hbar\omega_1$  and $n_2\hbar\omega_2$ in Tab.~\ref{table_full} are consistent with the potential energy surfaces obtained from DFT calculations shown in
Fig.~\ref{fig_overview}(a), which exhibit a highly asymmetric double well, with a shallow well for the down-state and a deep one for the up-state. The asymmetry
increases as the tip-surface separation is reduced. In particular, the depth of the potential well of the up-state ($n_1\hbar\omega_1$), which according to
Tab.~\ref{table_full} amounts to 0.53eV at 7.3\AA{}, agrees quite well with that calculated within DFT, whereas the model predicts a down-state well of 0.17eV at
7.3\AA{} that is slightly deeper than that derived from the \textit{ab-initio} calculations (cf. Fig.~\ref{fig_overview}(a)). This may be due to the fact that the
potentials in Fig.~\ref{fig_overview}(a) were calculated for a single PTCDA molecule, whereas in the switching experiments edge molecules were used; their
hydrogen bonds to neighboring molecules will lead to a significant increase of the barrier height. Note, however, that the model does correctly predict the
decrease in depth of the down-state well as the tip-surface separation is decreased; this tendency is due to the reduction of the potential minimum to a saddle
point for tip-surface separations of less than about 6.2\AA{} (cf. Fig.~\ref{fig_overview}(a)).

\begin{figure}
 \centering
 \includegraphics[width=0.55\linewidth,clip]{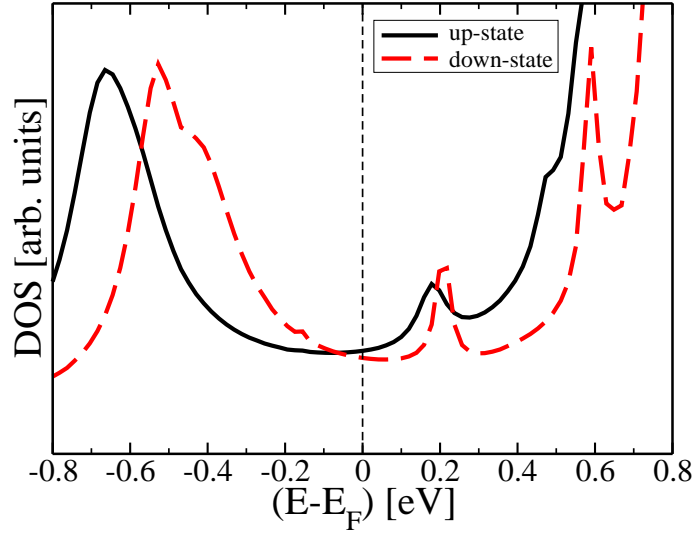}
 \caption{(color online) Density of states obtained from the DFT calculations described in Ref.~\citenum{Toher1} for the PTCDA molecule on the Ag(111) surface
          (down-state--dashed red) and attached to the tip (up-state--black) for a tip--surface separation of 7\AA{}.
          The level just above the Fermi energy is at the same position as in the simple model described here. The level below, however, is lower in energy
          compared to the model but also compared to the experiments.}
 \label{dos_dft}
\end{figure}

\begin{figure}
 \centering
 \includegraphics[width=0.55\linewidth,clip]{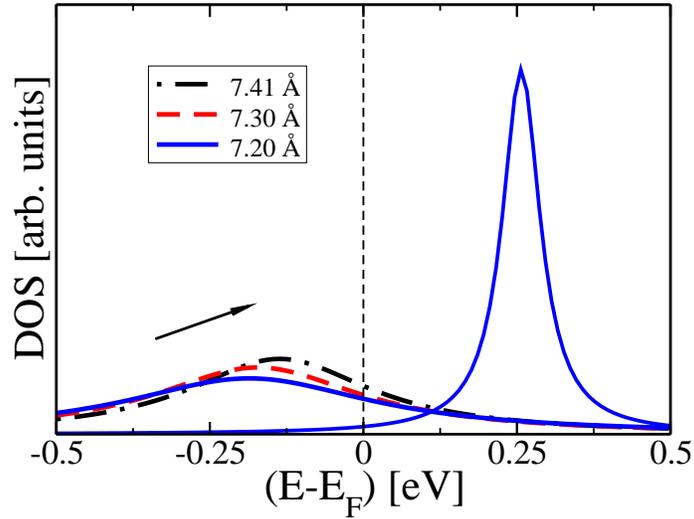}
 \caption{(color online) Density of states obtained from the fitting procedure for different tip--surface separations. If the molecule is attached to the tip
 the level below the Fermi energy moves up with increasing distances while the level above only shifts slightly.}
 \label{dos}
\end{figure}

A further important parameter in our model for the transfer rate is the position of energy level $\varepsilon_{m}$ through which the electron current that causes
the vibrational heating passes (i.e.~the transport level), because this influences the energy dependent density of states that enters the rate via
Eqs.~(\ref{rate_up}) and (\ref{rate_down}). It is clear that levels on either side and closest to the Fermi energy $\varepsilon_F$ are the most important channels for the electron
current. Our DFT calculations\cite{Toher1} show that mainly states both above and below the Fermi level could contribute, see Fig.~\ref{dos_dft}. The level below
$\varepsilon_F$ is the former LUMO that gets filled on adsorption and that is clearly observed in scanning tunneling spectroscopy.\cite{tautz07,tautz07a,Kraft}
The sharp level above $\varepsilon_F$ that is found in DFT appears in experiments as a broader feature in the gap between the former LUMO and the LUMO+1,
especially for molecules at the edges of monolayer islands. In our minimal model Hamiltonian we can only take one transport level into account. It turns out that
the qualitatively different behavior of the two processes surface$\rightarrow$tip and tip$\rightarrow$surface requires the use of two different transport levels,
depending on the switching direction. This is reflected in Tab.~\ref{table_full} by negative values $\varepsilon_{m,1}$ for the tip$\rightarrow$surface process,
while the surface$\rightarrow$tip process has positive $\varepsilon_{m,2}$ values (the spectral density of the levels $\varepsilon_{m,1}$ and $\varepsilon_{m,2}$
are shown in Fig.~\ref{dos}). In other words, we have to assume that in the up-state the switching current passes mainly through occupied DOS of the junction,
whereas in the down-state it passes predominantly through the empty DOS of the adsorbed molecule. Note that due to the way in which the bias voltage drops between
tip and substrate, both molecular levels $\varepsilon_{m,1}$ and $\varepsilon_{m,2}$ are within the bias window and may in principle contribute to the transport,
but in our minimal model we can -- as mentioned above -- only take one into account at a time. 

The fitted values $\varepsilon_{m,1}$ show a clear tendency to move up towards the Fermi level as the tip-surface separation is increased. This tendency is known 
very well both from experiment\cite{PTCDA_Kondo} and DFT calculations,\cite{Pump1,Toher1} although the precise level positions in experiment and \textit{ab initio}
theory differ from those in Tab.~\ref{table_full}. This is not too surprising since our minimal model only allows for a single Lorentzian level whereas the actual
density of states is much more complicated. The fitted values $\varepsilon_{m,2}$ range between 0.24eV and 0.30eV, whereas the DFT calculation has this level fixed
at 0.2eV.
 
The small decrease in the transfer rate of the surface$\rightarrow$tip process at about 220meV (Fig.~\ref{fit_pioa}(a)) is due to the molecular level
$\varepsilon_{m,2}$ entering into resonance with the Fermi energy of the STM tip, which leads to a reduction of the vibrational lifetimes of the PTCDA molecule in
the junction (\textit{i.e.} the rate $\Gamma_{\downarrow}$ at which the molecular vibrational energy dissipates into the electrodes is increased, cf. Eqs.~(A1, A4)).
This in turn reduces the transfer rate of the molecule between the surface and the tip. Further raising the bias voltage beyond this point results in the transfer
rate increasing once again due to non-resonant tunneling. We stress that this behavior can only be obtained if an energy-dependent DOS is used; a constant DOS could
not yield such a behavior. Unfortunately, the increase above 240meV cannot be observed in the experiments since the molecule normally disintegrates at lower biases
than this because of the high current density.

\begin{table}
 \centering
 \caption{Model-parameters for the switching of the PTCDA between the tip to the surface obtained by fitting $R$ to the experiments.
 The subscripts ``1'' and ``2'' indicate switching from STM tip to the surface or the reverse process, respectively.
 Energies $\varepsilon_{m}$, $\hbar \omega$, $n \hbar \omega$, $\Delta_s$ and $\Delta_t$ are given in meV.
 The dimensionless parameters $\lambda_{1,2}\:=\:\lambda_0/\hbar\omega_{1,2}$ describe the electron-vibration interaction.}
 \label{table_full}
 \begin{tabular}{cccccccc}
  \hline
  \hline
   Tip--surface distance & $\lambda_1$ & $\varepsilon_{m,1}$ & $n_1$ & $\hbar \omega_1$ & $n_1 \hbar \omega_1$ & $\Delta_{s,1}$ & $\Delta_{t,1}$\\
  \hline
   $7.17$\AA & $0.025$ & $-187$ & $17$ & $40.68$ & $692$ & $155$ & $80.00$\\
   $7.20$\AA & $0.025$ & $-186$ & $16$ & $40.71$ & $651$ & $165$ & $80.00$\\
   $7.24$\AA & $0.025$ & $-186$ & $15$ & $40.76$ & $611$ & $168$ & $75.02$\\
   $7.27$\AA & $0.025$ & $-186$ & $14$ & $40.78$ & $571$ & $148$ & $66.23$\\
   $7.30$\AA & $0.025$ & $-172$ & $13$ & $40.82$ & $531$ & $148$ & $56.92$\\
   $7.34$\AA & $0.025$ & $-141$ & $12$ & $40.86$ & $490$ & $148$ & $44.80$\\
   $7.37$\AA & $0.025$ & $-139$ & $11$ & $40.89$ & $450$ & $148$ & $39.57$\\
   $7.41$\AA & $0.025$ & $-136$ & $10$ & $40.94$ & $409$ & $148$ & $33.59$\\
   $7.44$\AA & $0.024$ & $-124$ & $9$ & $40.98$ & $369$ & $148$ & $26.30$\\
   $7.47$\AA & $0.010$ & $-110$ & $8$ & $41.00$ & $328$ & $148$ & $26.87$\\
  \\
  \hline
  \hline
   Tip--surface distance
 & $\lambda_2$ & $\varepsilon_{m,2}$ & $n_2$ & $\hbar \omega_2$ & $n_2 \hbar \omega_2$ & $\Delta_{s,2}$ & $\Delta_{t,2}$\\
  \hline
   $7.17$\AA & $0.012$ & $260$ & $8$ & $18.95$ & $152$ & $24$ & $11.9$\\
   $7.20$\AA & $0.009$ & $257$ & $8$ & $19.00$ & $152$ & $25$ & $12.5$\\
   $7.24$\AA & $0.007$ & $253$ & $9$ & $19.06$ & $172$ & $30$ & $14.9$\\
   $7.27$\AA & $0.006$ & $246$ & $9$ & $19.12$ & $172$ & $28$ & $13.9$\\
   $7.30$\AA & $0.005$ & $260$ & $9$ & $19.15$ & $172$ & $24$ & $11.8$\\
   $7.34$\AA & $0.004$ & $259$ & $9$ & $19.22$ & $173$ & $22$ & $10.8$\\
   $7.37$\AA & $0.003$ & $269$ & $10$ & $19.26$ & $193$ & $23$ & $11.4$\\
   $7.41$\AA & $0.002$ & $285$ & $11$ & $19.32$ & $213$ & $23$ & $10.1$\\
   $7.44$\AA & $0.002$ & $299$ & $10$ & $19.37$ & $194$ & $10$ & $4.9$\\
   $7.47$\AA & $0.0004$ & $258$ & $10$ & $19.42$ & $194$ & $28$ & $8.9$\\
  \hline
 \end{tabular}
\end{table}

\section{conclusion}
In summary, switching between low- and high conductance states has been observed in a single molecule junction consisting of a PTCDA molecule on a Ag(111)
substrate and contacted by an STM tip. The rates for the transition between these two states can be sensitively tuned by varying the applied bias as well as the
tip-surface separation. A vibrational heating mechanism where molecular bonds are excited by tunneling charges has been proposed to interpret the experimental
results. Switching rates were calculated within a minimal model Hamiltonian approach describing the interaction between tunneling electrons and local molecular
vibrations. The experimental results could be fitted over a broad voltage range for the cases where the PTCDA molecule switches both from the surface to the tip
and from the tip to the surface. In particular, the non-monotonic behavior of the surface to tip switching rate could only be described by modeling the
DOS by Lorentzian functions instead of assuming it to be energy independent, as has been the usual practice in the literature until now.
This demonstrates that it is crucial to take the non-constant behavior of the molecular DOS into account.

\begin{acknowledgments}
This work has been supported by the German Priority Program ``Quantum transport at the molecular scale (SPP1243)''.
The authors acknowledge the Center for Information Services and High Performance Computing (ZIH) at the Dresden University of Technology for computational
resources. GC acknowledges the South Korean Ministry of Education, Science, and Technology Program, Project WCU ITCE No. R31-2008-000-10100-0.
TB would like to acknowledge an especially fruitful discussion with Florian Pump.
\end{acknowledgments}

\appendix

\section{Calculation of the transition rates}
In the following we want to sketch the derivation of the transition and relaxation rates. The terms $\Gamma_{\uparrow,\downarrow}^{ss}$ and
$\Gamma_{\uparrow,\downarrow}^{tt}$ are all similar, and it is sufficient to calculate explicitly only the term $\Gamma_{\downarrow}^{ss}$. Inserting the
electron-vibration interaction (Eq.~(\ref{h_ev})) into Eq.~(\ref{rate_down}) together with the expression for the molecular DOS gives
\begin{align}
 \Gamma_\downarrow^{ss}=&\:2\frac{\pi\left(\varepsilon_m'\right)^2}{M\,\omega}
                         \:\sum_{\alpha',\alpha}\:\left|\braket{\alpha'}{m}\braket{m}{\alpha}\right|^2
                           \left[1-f_s(\varepsilon_{\alpha'})\right]
                         \:f_s(\varepsilon_\alpha)\:\delta(\varepsilon_{\alpha'}-\varepsilon_\alpha-\hbar\omega)\notag\\
 =&\:2\frac{\pi\left(\varepsilon_m'\right)^2}{M\,\omega}\: \int d\varepsilon\, \rho_m^s(\varepsilon)\, \rho_m^s(\varepsilon + \hbar\omega)
      \left[1 - f_s(\varepsilon + \hbar\omega)\right]\, f_s(\varepsilon)\notag\\
 =&\:2\frac{\Delta_s^2\,\left(\varepsilon_m'\right)^2}{M\,\omega\,\pi}\:\int d\varepsilon
    \,\frac{1}{\left[\varepsilon-\varepsilon_m\right]^2\:+\:\Delta^2}
      \frac{1}{\left[\varepsilon+\hbar\omega-\varepsilon_m\right]^2\:+\:\Delta^2}
    \left[1 - f_s(\varepsilon + \hbar\omega)\right]\, f_s(\varepsilon)\notag\\
 \approx&\:2\frac{\Delta_s^2\,\left(\varepsilon_m'\right)^2}{M\,\omega\,\pi}\:\int d\varepsilon
    \,\frac{1}{\left[\varepsilon-\varepsilon_m\right]^2\:+\:\Delta^2}
      \frac{1}{\left[\varepsilon+\hbar\omega-\varepsilon_m\right]^2\:+\:\Delta^2}\notag\\
  &\:\times\left[1 - \Theta(\varepsilon_{Fs}-\hbar\omega-\varepsilon)\right]\, \Theta(\varepsilon_{Fs}-\varepsilon)\notag\\
 =&\:2\frac{\Delta_s^2\,\left(\varepsilon_m'\right)^2}{M\,\omega\,\pi}\:\int\limits_{\varepsilon_{Fs}-\hbar\omega}^{\varepsilon_{Fs}} d\varepsilon
    \,\frac{1}{\left[\varepsilon-\varepsilon_m\right]^2\:+\:\Delta^2}
      \frac{1}{\left[\varepsilon+\hbar\omega-\varepsilon_m\right]^2\:+\:\Delta^2}\notag\\
 =&\:\frac{4\,\Delta_s^2\,\lambda_0^2}{\pi\,\Delta\,\hbar^2\omega\,\left(4\,\Delta^2\,+\,\hbar^2\omega^2\right)}
   \Biggl\{\hbar\,\omega\left(\tan^{-1}\left[\frac{\varepsilon_m-\varepsilon_{Fs}+\hbar\omega}{\Delta}\right]\right.\Biggr.\notag\\
  &\:\left.-\tan^{-1}\left[\frac{\varepsilon_m-\varepsilon_{Fs}-\hbar\omega}{\Delta}\right]\right)
   \:+\:\Delta\left(\log\left[\Delta^2+\left(\varepsilon_m-\varepsilon_{Fs}+\hbar\omega\right)^2\right]\right.\notag\\
\label{gamma_ss_d}
  &\:\Biggl.\left.+\log\left[\Delta^2+\left(\varepsilon_m-\varepsilon_{Fs}-\hbar\omega\right)^2\right]
                -2\,\log\left[\Delta^2+\left(\varepsilon_m-\varepsilon_{Fs}\right)^2\right]\right)\Biggr\}
\end{align}
In the first step the sum over states has been replaced with an integral over $\varepsilon$ by introducing $\rho_m^s(\varepsilon)$. In the second step
the expression for the molecular DOS was used to rewrite the local density of states. Since the STM experiments are carried out at 5-6 K one can approximate
the Fermi function with the Heaviside step function in the next step. Thus, the limits of the integral can be changed from $+\infty$ and $-\infty$ to
$\varepsilon_{Fs}$ or $\varepsilon_{Fs}-\hbar\omega$ respectively. The influence of an applied bias can be easily introduced by shifting the Fermi level of
the surface $\varepsilon_{Fs}=\varepsilon_{F0s}+eV$, where $\varepsilon_{F0s}$ is the Fermi level at $V=0$ of the surface. Since we used the low temperature
approximation in step 3 of Eq.~(A1) the excitation rates $\Gamma_{\uparrow}^{ss,tt}$ become zero, because of the Pauli exclusion principle. The Pauli
exclusion principle also simplifies the calculation of the remaining terms $\Gamma_{\uparrow,\downarrow}^{ts}$ and $\Gamma_{\uparrow,\downarrow}^{st}$, which
describe the transition rates due to the inelastic scattering of tunneling electrons between surface and tip. Assuming
$\varepsilon_{Ft}=\varepsilon_{Fs}:=\varepsilon_{F}$, for positive applied bias the tunneling from surface to tip through the adsorbate level is prohibited.
The excitation is forbidden because all states at the tip are occupied up to the energy $\varepsilon_F$, thus making it impossible for an electron from the
surface with energy $\varepsilon_F-|eV|-\hbar\omega$ to tunnel to the tip. The probability of relaxing an adsorbate vibrations due to the inelastic scattering
of tunneling electrons from the surface to the tip is negligibly small, because the scattered electron would need several $\hbar\omega$ to gain enough energy.
But this process can also be excluded, since the electron-vibration interaction on the adsorbate vibration is in general weak and we treat it by first-order
perturbation theory. Thus, the transition rates can be written as, \textit{e.g.}
\begin{align}
\label{cases_up_rates}
 \Gamma_\uparrow^{st}=\quad&\begin{cases}2\frac{\pi\left(\varepsilon_m'\right)^2}{M\,\omega}\:
                                       \int\limits_{\varepsilon_{F}-\left|eV\right|+\hbar\omega}
                                                  ^{\varepsilon_{F}} d\varepsilon\;\;
                                           \rho_m^{s}(\varepsilon-\hbar\omega)\,
                                           \rho_m^{t}(\varepsilon)
                                        &\forall\:\left|eV\right|>\hbar\omega\\ \\
                                       0\quad&\forall\:\left|eV\right|\leq\hbar\omega\\ \\\end{cases}
\end{align}
\begin{align}
  \Gamma_\uparrow^{st}\;\overset{\left|eV\right|>\hbar\omega}{=}&\;\frac{4\,\Delta_s\,\Delta_t\,\lambda_0^2}
                                                                        {\pi\,\Delta\,\hbar^2\omega\,\left(4\Delta^2+\hbar^2\omega^2\right)}\,
                                            \Biggl\{\hbar\omega\,\left(\tan^{-1}\left[\frac{-\varepsilon_m+\varepsilon_F}{\Delta}\right]
                                                                      +\tan^{-1}\left[\frac{\varepsilon_m-\varepsilon_F+\left|eV\right|}{\Delta}\right]\right.\Biggr.\notag\\
 &\;                                                            \left.+\tan^{-1}\left[\frac{\varepsilon_m-\varepsilon_F-\hbar\omega+\left|eV\right|}{\Delta}\right]
                                                                      +\tan^{-1}\left[\frac{-\varepsilon_m+\varepsilon_F-\hbar\omega}{\Delta}\right]\right)\notag\\
 &\;                                                    +\Delta\,\left(\log\left[\Delta^2+\left(\varepsilon_m-\varepsilon_F\right)^2\right]
                                                                      +\log\left[\Delta^2+\left(\varepsilon_m-\varepsilon_F+\left|eV\right|\right)^2\right]\right.\notag\\
\label{gamma_st_u}
 &\;\Biggl.\left.                                                     -\log\left[\Delta^2+\left(\varepsilon_m-\varepsilon_F-\hbar\omega+\left|eV\right|\right)^2\right]
                                                                      -\log\left[\Delta^2+\left(\varepsilon_m-\varepsilon_F+\hbar\omega\right)^2\right]
                                                           \right)
                                            \Biggr\} \quad,
\end{align}
\begin{align}
 \Gamma_\downarrow^{st}=&\:2\frac{\pi\left(\varepsilon_m'\right)^2}{M\,\omega}
                         \:\int\limits_{\varepsilon_{F}-\left|eV\right|-\hbar\omega}^{\varepsilon_{F}} d\varepsilon\;\;
                                                                                       \rho_m^{s}(\varepsilon + \hbar\omega)\, \rho_m^{t}(\varepsilon)\notag\\\notag\\
 =&\:\frac{4\,\Delta_s\,\Delta_t\,\lambda_0^2}
          {\pi\,\Delta\,\hbar^2\omega\,\left(4\Delta^2+\hbar^2\omega^2\right)}
    \:\Biggl\{\hbar\omega\,\left(\tan^{-1}\left[\frac{\varepsilon_m-\varepsilon_F+\left|eV\right|}{\Delta}\right]
                                       +\tan^{-1}\left[\frac{-\varepsilon_m+\varepsilon_F+\hbar\omega}{\Delta}\right]\right.\Biggr.\notag\\
  &\:\left.                            +\tan^{-1}\left[\frac{\varepsilon_m-\varepsilon_F+\hbar\omega+\left|eV\right|}{\Delta}\right]
                                       +\tan^{-1}\left[\frac{-\varepsilon_m+\varepsilon_F}{\Delta}\right]\right)\notag\\
  &\:                    +\Delta\,\left(\log\left[\Delta^2+(\varepsilon_m-\varepsilon_F-\hbar\omega)^2\right]
                                       +\log\left[\Delta^2+(\varepsilon_m-\varepsilon_F+\hbar\omega+\left|eV\right|)^2\right]\right.\notag\\
\label{gamma_st_d}
  &\:\Biggl.\left.                     -\log\left[\Delta^2+(\varepsilon_m-\varepsilon_F)^2\right]
                                       -\log\left[\Delta^2+(\varepsilon_m-\varepsilon_F+\left|eV\right|)^2\right]\right)\Biggr\}
\quad.
\end{align}
The parameter $\lambda_0$ given in Eq.~(\ref{lambda0}) is an important parameter in our theory, as one can clearly see in the Eqs.~(A1--A4). In contrast to
all other parameters, \textit{e.g.} the broadening $\Delta$ or the energy $\varepsilon_m$, it is in general difficult to determine it from experiment or
\textit{ab-initio} calculations. However, these are only prefactors which change the absolute magnitude of the transition rates and thus can be easily fitted
to the experiments.

\end{document}